\documentstyle[aps,epsfig,preprint]{revtex}

\newcommand{\be}{\begin{equation}}
\newcommand{\ee}{\end{equation}}
\newcommand{\ba}{\begin{eqnarray}}
\newcommand{\ea}{\end{eqnarray}}
\begin{document}
\draft

\title{
Fluctuations  and Deconfinement Phase Transition \\
in Nucleus--Nucleus Collisions }

\author{
{\bf M. Ga\'zdzicki}$^{a,b}$, {\bf M.I. Gorenstein}$^{c,d}$ and {\bf St.
Mr\'owczy\'nski}$^{e,b}$ }

\address{
$^a$ Institut f\"ur  Kernphysik, Universit\"at  Frankfurt, Germany\\
$^b$ Institute of Physics, \'Swi\c etokrzyska Academy, Kielce, Poland \\
$^c$ Bogolyubov Institute for Theoretical Physics, Kiev, Ukraine\\
$^{d}$ Institut f\"ur Theoretische Physik, Universit\"at Frankfurt,
Germany\\
$^e$ So\l tan Institute for Nuclear Studies, Warsaw, Poland\\
}

\date{\today}

\maketitle

\begin{abstract}
\noindent
We propose a method to experimentally study  the equation of state of
strongly interacting matter created at the early stage of nucleus--nucleus
collisions. The method exploits the relation between relative entropy and
energy fluctuations and equation of state. As a measurable quantity, the
ratio of properly filtered multiplicity to energy fluctuations is proposed.
Within a statistical approach to the early stage of nucleus-nucleus
collisions, the fluctuation ratio manifests a non--monotonic collision
energy dependence with a maximum in the domain where the onset of
deconfinement occurs.

\end{abstract}

\newpage


\vspace{0.2cm} \noindent {\bf 1.} Nucleus--nucleus (A+A) collisions at
high energies provide a unique opportunity to study properties of strongly
interacting matter which at sufficiently high energy density is predicted
to exist in a deconfined or quark-gluon-plasma phase. Success of the
statistical models to strong interactions \cite{Braun-Munzinger:2003zd}
suggests that the system created in these collisions is close to
thermodynamical equilibrium. Consequently, the properties of the matter
are naturally expressed in terms of its equation of state (EoS) which
in turn is sensitive to possible phase transitions. Increasing the energy
of nuclear collisions, one expects to achieve at the collision early stage
higher and higher energy density that at a certain point is sufficient
for creation of the quark-gluon plasma. Then, EoS should experience
a qualitative change. Observing a clear signal of this change is among
main tasks of the whole experimental program of study A+A collisions.
The task, however, has appeared rather difficult. It is far not simple
to express thermodynamical characteristics at the early stage through
the directly measurable quantities. The entropy is of particular interest,
as it is believed to be conserved during the expansion of the matter,
and several methods to determine it experimentally have been suggested
\cite{Csernai:qf,Bialas:1999wi,Gazdzicki:ze}.  Other observables, which
may be  sensitive to the EoS of the early stage matter, have been also
proposed.  Transverse momentum spectra \cite{VanHove:1982vk}, two pion
correlations \cite{Hung:1994eq}, anisotropic flow \cite{Ollitrault:bk}
and strangeness production \cite{Gazdzicki:1998vd} are discussed in this
context.

\vspace{0.2cm} \noindent {\bf 2.}
The recently measured energy dependence of the pion multiplicity,
which is related to the system's entropy, and kaon (system's strangeness)
production in central Pb+Pb collisions \cite{Afanasiev:2002mx,Alt:2003rn}
show the changes which are consistent with the hypothesis
\cite{Gazdzicki:1998vd,Gorenstein:2003cu} that a transient state of
deconfined matter is created at the collision energies higher than
about 30 A$\cdot$GeV in fixed target experiments. This conclusion is
reached within the Statistical Model of the Early Stage, SMES
\cite{Gazdzicki:1998vd}, which assumes creation of the matter
(in confined, mixed or deconfined phase) at early stage of the
collision according to the maximum entropy principle.

\vspace{0.2cm} \noindent {\bf 3.} In this letter we propose a new method
of study of EoS which uses the ratio of properly filtered multiplicity
and energy fluctuations as directly measurable quantity and refers to
SMES \cite{Gazdzicki:1998vd} as a physical framework. Within this model
the ratio is directly related to the fluctuations of the early  stage
entropy and energy and thus is sensitive to the EoS of the early stage
matter. We show here that the model predicts a non-monotonic energy
dependence of the ratio with the maximum where the onset of deconfinement
occurs.

\vspace{0.2cm} \noindent {\bf 4.} In thermodynamics, the energy $E$,
volume $V$ and entropy $S$ are related to each other through EoS. Thus,
various values of the energy of the initial equilibrium state lead to
different, but uniquely determined, initial entropies. When the collision
energy is fixed the energy, which is used for particle production, still
fluctuates. These fluctuations of the inelastic energy are caused by the
fluctuations in the dynamical process which leads to the particle
production. They are called here the {\it dynamical} energy fluctuations.
Clearly, the {\it dynamical} energy fluctuations lead to the {\it
dynamical} fluctuations of entropy, and the relation between them is, in
the thermodynamical approach, given by EoS. Consequently, simultaneous
event--by--event measurements of both the entropy and energy should yield
an information on EoS. Since EoS manifests an anomalous behavior in a
phase transition region the anomaly should be also visible in the ratio of
entropy to energy fluctuations.

\vspace{0.2cm} \noindent {\bf 5.}
The energy and entropy can be defined in any form of matter, confined,
mixed  and deconfined, in the collision early stage and in the system's
final state. If the produced matter can be treated as an isolated system,
the energy is obviously conserved. The entropy is also expected to be
conserved during the system's expansion and freeze--out. However, there
is a significant difference between the two quantities. While the energy
is defined for every event the entropy refers to an ensemble of events.

\vspace{0.2cm} \noindent {\bf 6.}
Since we are going to discuss the collision energy dependence of the
fluctuations within the SMES \cite{Gazdzicki:1998vd}, let us present
the model's basic assumptions. The volume, $V$, where the matter in
confined, mixed or deconfined state is produced at the collision early
stage, is given by the Lorentz contracted volume occupied by wounded
nucleons. For the most central collisions the number of wounded
nucleons $N_W \approx 2 \: A$. The net baryonic number of the {\em
created} matter equals zero. Even in the most central A+A collisions,
only a fraction of the total collision energy is used for a particle
production. The rest is taken away by the baryons which contribute to the
baryon net number.

\vspace{0.2cm} \noindent {\bf 7.} The fluctuations occurring in the
collision early stage, which are local in coordinate or momentum space,
are washed out, at least partially, in the course of temporal evolution of
the fireball due to relaxation processes such as particle diffusion, see
e.g. \cite{Shuryak:2000pd}. This probably explains why the electric
charge fluctuations generated at the QGP phase 
\cite{Asakawa:2000wh,Jeon:2000wg}, which are significantly smaller
than those in the hadron phase, are not seen in the experimental data
\cite{Adcox:2002mm,Adams:2003st,Blume:2002mr}. It should be stressed, 
however, that the relaxation processes are irrelevant for 
our considerations as we are interested in the fluctuations of 
{\em total} inelastic energy and entropy of the system created at the collision 
early stage. 
Because of the exact energy and approximate entropy 
conservation the fluctuations observed in the final state equal to the 
early stage fluctuations. We assume here that all produced particles 
are detected but further we relax this assumption.  The 
inelastic energy deposited 
in the fireball for the particle production should not be confused with 
the collision energy. While the former one fluctuates  
the latter is fixed and it does not fluctuate at all.

\vspace{0.2cm} \noindent {\bf 8.}
We denote by $\delta E$ the event--by--event deviations of the energy from
its average value $E$ caused by the dynamical fluctuations which occur in
the thermalization process. We assume that $\delta E \ll E$.
As $E = \varepsilon \, V$, where $\varepsilon$ is the energy density.
One has $\delta E=V\delta \varepsilon + \varepsilon~ \delta V~,$
i.e. the change of the system's energy is due to the changes of the system's
energy density and volume which are considered further as two independent
thermodynamical variables. The energy density is usually a unique function
of the temperature, $T$, but when the system experiences a first order
phase transition, $\varepsilon$ in the mixed phase depends on the
relative abundance of each phase.

\vspace{0.2cm} \noindent {\bf 9.}
According to the first and the second
principles of thermodynamics, the entropy change $\delta S$ is given as $T
\delta S = \delta E + p \delta V$, which provides
$T\delta S = V \delta \varepsilon + (p + \varepsilon ) \delta V$,
where $p$ is the pressure. Using the identity $TS = E + pV$ one finds
\begin{equation}\label{therm1}
\frac{\delta S}{S} = \frac{1}{1 +p/\varepsilon} \: \frac{\delta
\varepsilon}{\varepsilon} + \frac{\delta V}{V} \;.
\end{equation}

\vspace{0.2cm} \noindent {\bf 10.}
When $\delta \varepsilon=0$, i.e. when the fluctuations of the initial
energy and entropy are entirely due to the volume fluctuations at
a constant energy density, Eq.~(\ref{therm1}) provides:
$\delta S/S=\delta V/V= \delta E/E$.
Thus, the relative dynamical fluctuations of entropy are exactly equal to
those of energy and they are insensitive to the form of EoS. The $\delta
\varepsilon=0$ limit may serve as an approximation for all inelastic A+A
collisions where fluctuations of the collision geometry dominate all other
fluctuations. This case, however, is not interesting from our point of
view.

\vspace{0.2cm} \noindent {\bf 11.}
When $\delta V=0$ the fluctuations of the initial energy, $\delta E$,
are entirely due to the energy density fluctuations. In this case
Eq.~(\ref{therm1}) gives:
\begin{equation}\label{T}
\frac{\delta S}{S} ~= ~\frac{\delta E}{E} \: \frac{1}{1 +
p/\varepsilon} \;.
\end{equation}
As seen, $\delta S/S$ is now sensitive, via the factor
$(1+p/\varepsilon)^{-1}$, to the EoS at the early stage of A+A collision.
We are interested just in such a situation.

\vspace{0.2cm} \noindent {\bf 12.} The number of wounded nucleons can, in
principle, be measured on the event--by--event basis. This can be achieved
by measuring the number of spectator nucleons, $N_{S}$, in the so-called
zero degree calorimeter, used in many experiments. Then, $N_W \approx 2 (A
- N_{S})$. Selecting the most central events, we can neglect contribution
from the impact parameter variation. Since the system's volume, as defined
in SMES, is then fixed the entropy fluctuations are given by
Eq.~(\ref{T}).

\vspace{0.2cm} \noindent {\bf 13.}
To study the entropy fluctuations it appears convenient to introduce the
ratio of relative fluctuations:
\begin{equation}\label{R}
R_e \equiv \frac{(\delta S)^2/S^2}{(\delta E)^2/E^2} =
\left(1+\frac{p}{\varepsilon}\right)^{-2} \;,
\end{equation}
which qualitatively behaves as follows. The ratio $p/\varepsilon$ is about
$1/3$ in both the confined phase and in the hot quark-gluon plasma (QGP).
Then, $R_e \approx (3/4)^{2}\cong 0.56$ and it is rather independent of
the collision energy except the domain where the initially created matter
experiences the deconfinement phase transition. An exact nature of the
transition is unknown but modelling of the transition by means of the
lattice QCD \cite{Karsch:2001cy} shows a very rapid change of the
$p/\varepsilon$ ratio in a narrow temperature interval $\Delta T \cong
5$~MeV where the energy density grows by about an order of magnitude
whereas the pressure remains nearly unchanged. One refers to this
temperature interval as a `generalized mixed phase'. The ratio
$p/\varepsilon$ reaches minimum at the so-called softest point of the EoS
\cite{Hung:1994eq} which corresponds to a maximum of $R_e \approx 1$.
Consequently, we expect a non--monotonic behavior of the ratio $R_e$ as a
function of the collision energy.

\vspace{0.2cm} \noindent {\bf 14.} The energy dependence of the
fluctuation ratio $R_e$ calculated within SMES \cite{Gazdzicki:1998vd}
(using its standard values of all parameters) is shown in Fig. 1. We
repeat here that the model correctly reproduces the energy dependence of
pion and strangeness production and it relates experimentally observed
anomalies to the onset of deconfinement. Within the model, the confined
matter, which is modelled as an ideal gas, is created at the collision
early stage below the energy of 30 A$\cdot$GeV. In this domain, the ratio
$R_e$ is approximately independent of collision energy and equals about
0.6. The model assumes that the deconfinement phase-transition is of the
first order. Thus, there is the mixed phase region, corresponding to the
energy interval 35$\div$60 A$\cdot$GeV, where $R_e$ ratio increases and
reaches its maximum, $R_{e}\approx 0.8$, at the end of the transition
domain. Further on, in the pure QGP phase represented by an ideal
quark-gluon gas under bag pressure, the ratio decreases and $R_e$
approaches its asymptotic value 0.56 at the highest SPS energy
160A$\cdot$GeV. Small deviations from $p=\varepsilon/3$ are in SMES due to
non-zero masses of strange degrees of freedom, both in confined and
deconfined phases, and due to the bag pressure in QGP. The two effects can
be safely neglected at $T \gg T_{c}$.

\vspace{0.2cm} \noindent {\bf 15.}
In principle, the initial energy fluctuations might be sizable while our
analysis holds for infinitesimally small fluctuations as the ratio
$R_e$ (\ref{R}) is defined above by introducing the dynamical energy
fluctuations $\delta E$ and we use thermodynamical identities to calculate
the entropy fluctuations $\delta S$. However, the calculations with explicit
initial energy distribution show that the finite size of initial energy
fluctuations does not much change our results. The dependence of $R_e$
on the collision energy shown in Fig.~1 remains essentially the same.
The only difference is a `smooth' behavior of $R_e(F)$ near the maximum.

\vspace{0.2cm} \noindent {\bf 16.}
The early stage energy and entropy fluctuations are not directly observable,
however, as we discuss in the remaining part of the paper, $R_e$ can be
inferred from the experimentally accessible information. Since the energy
of an isolated system is a conserved quantity, one measures the initial
energy deposited for the particle production, summing up the final state
energies of {\em all} produced particles. The system's entropy is not
strictly conserved but, as already discussed, it is approximately conserved.
Therefore, the final state entropy of {\em all} produced particles is close
to the initial entropy. The entropy cannot be directly measured but
it can be expressed through measurable quantities.

\vspace{0.2cm} \noindent {\bf 17.}
As well known, the system's entropy is related to the mean particle
multiplicity. For example, $\overline{N} = S/3.6$ in the ideal gas of
massless bosons. The relation is, in general, more complex but we assume
that the final state mean multiplicity is proportional to the initial state
entropy, i.e. $\overline{N} \sim S$. With the over-bar we denote averaging
over events that have identical initial conditions (the same amount of energy
is deposited for the particle production). It is clear that for the class of
events with a  fixed value of $\overline{N}$, the multiplicity $N$ measured
in each event fluctuates around $\overline{N}$. These are
{\em statistical} but not dynamical fluctuations. We note that particle
multiplicity can be determined for every event, in contrast to the entropy
which is defined by averaging of hadron multiplicities in the ensemble of
events. Since $\overline{N} \sim S$, we get: $ \delta S/S = \delta
\overline{N}/\overline{N}$. Thus, the dynamical entropy fluctuations
are equal to the dynamical fluctuations of the mean multiplicity. It is
crucial to distinguish the dynamical fluctuations of $\overline{N}$ from
the statistical fluctuations of $N$ around $\overline{N}$. We clarify this
point below.

\vspace{0.2cm} \noindent {\bf 18.}
The multiplicity $N$ measured on event--by--event basis varies not only
due to the dynamical fluctuations at a collision early stage but
predominately due to the statistical fluctuations at freeze--out. Thus,
the final multiplicity distribution, ${\cal P}(N)$, is given by:
\begin{equation}\label{PN}
{\cal P}(N) = \int_{0}^{\infty}d\overline{N}
~W(\overline{N})~P_{\overline{N}}(N) ~,
\end{equation}
where $W(\overline{N})$ describes fluctuations of $\overline{N}$ due to
dynamical fluctuations of $E$, and $P_{\overline{N}}(N)$ is the
statistical probability distribution of $N$ for a given $\overline{N}$.
The finally measured mean value of an observable $f(N)$ results from
averaging over the $W$ and $P$ distributions as
\begin{equation}\label{tot}
\langle \langle f_{N} \rangle \rangle \equiv \sum_{N} f(N) {\cal P}(N)=
\int_{0}^{\infty}d\overline{N}~ W(\overline{N})~ \sum_{N} f(N)
P_{\overline{N}}(N) \equiv \langle~ \overline{f(N)} ~\rangle \;.
\end{equation}
Thus, the complete averaging, $\langle \langle \cdots \rangle \rangle$, is
done in two steps: first -- the statistical, $ \overline{\cdots}
 \equiv \sum_{N}\cdots P_{\overline{N}}(N)$, and second
-- the dynamical averaging, $\langle \cdots \rangle \equiv
\int_{0}^{\infty}d\overline{N}\cdots W(\overline{N})$, one after another.
One easily shows that
\begin{equation}
\langle \langle N \rangle \rangle = \langle \overline{N} \rangle
~, ~~~~~~~
(\Delta N)^2 \equiv \langle \langle N^2 \rangle \rangle - \langle \langle
N \rangle \rangle^2 = (\delta \overline{N})^2
+ \langle (\delta N )^2 \rangle \;, \label{totvar}
\end{equation}
where $(\delta \overline{N})^2 \equiv \langle \overline{N}^2 \rangle -
\langle \overline{N} \rangle^2$ and $(\delta N)^2 \equiv \overline{N^2}
- \overline{N}^2$. Thus, the total fluctuations $(\Delta N)^{2}$, which
are experimentally measured, are
equal to the sum of the dynamical (early stage) fluctuations
$(\delta \overline{N})^2$ and the dynamically averaged statistical
fluctuations $\langle (\delta N)^{2}\rangle$ at freeze--out.

\vspace{0.2cm} \noindent {\bf 19.} We have considered above the ideal
detector which measures all produced particles. A real detector, however,
measures only a fraction of them, say charged particles in the limited
momentum acceptance of the detector. Let us denote the mean energy and
multiplicity of accepted particles as $\overline{E}_{A}$ and
$\overline{N}_{A}$. We assume that
\begin{equation}\label{dyn}
\frac{\delta \overline{E}_{A}}{\overline{E}_{A}}~=~ \frac{\delta
E}{E}~,~~~~\frac{\delta \overline{N}_{A}}{\overline{N}_{A}}~=~
\frac{\delta S}{S}~,
\end{equation}
i.e. relative dynamical fluctuations of the mean energy and mean
multiplicity of accepted particles are equal to the relative dynamical
fluctuations of the total energy and entropy in the initial state.
In our further considerations, we will omit the index `$A$', however,
it is understood that we deal with the accepted particles.

\vspace{0.2cm} \noindent {\bf 20.} There is a simple procedure to eliminate
the statistical fluctuations, and thus, to extract the dynamical
fluctuations of interest from the measured fluctuations, if
$P_{\overline{N}}(N)$ is the Poisson distribution.  Then, $(\delta
N)^{2}=\overline{N}$, and $(\delta \overline{N})^2 = (\Delta N)^2 -
\langle \langle N \rangle \rangle$. Therefore, the relative dynamical
fluctuations are expressed through the total relative fluctuations as
\be
\bigg({\delta \overline{N} \over \langle \langle N \rangle \rangle}
\bigg)^2 = \bigg({\Delta N \over \langle \langle N \rangle \rangle}
\bigg)^2 - {1 \over \langle \langle N \rangle \rangle} \;.
\ee
The distribution of energy $E$ of the system of several particles
is assumed to be of the form
\be
{\cal P}(E) =  \sum_N \int d\zeta \; W(\zeta) \; P_\zeta(N) \;
\int d\omega_1 \: P_\zeta(\omega_1) \;
\cdots \int d\omega_N \: P_\zeta(\omega_N) \;
\delta (E - \sum_{i=1}^{N}\omega_i) \;,
\ee
where $W(\zeta)$ describes dynamical fluctuations of the parameter
$\zeta$ which controls the multiplicity and energy fluctuations. In
principle,  $\zeta$ can be understood as a whole set of parameters.
$P_\zeta(N)$ is the multiplicity and $P_\zeta(\omega)$ single particle
energy distribution, both giving the statistical fluctuations.
One easily finds that
\ba
\langle \langle E \rangle \rangle
& = &
\langle \, \overline{N} \, \overline{\omega} \, \rangle  \;,
\\
(\Delta E)^2 &\equiv &
\langle \langle E^2 \rangle \rangle
- \langle \langle E \rangle \rangle^2
= (\delta \overline{E})^2
+ \langle (\delta E)^2 \rangle \;,
\ea
where $\overline{\omega^n} \equiv \int d\omega \; \omega^n \; P_\zeta
(\omega) $ and
\ba
(\delta \overline{E})^2 &\equiv&
\langle \, \overline{E}^2 \, \rangle - \langle \,\overline{E} \,\rangle^2
= \langle (\overline{N}\: \overline{\omega})^2 \rangle
- \langle \,\overline{N}\: \overline{\omega}\,\rangle^2  \;, \\
\langle (\delta E)^2 \rangle &\equiv&
\langle \, \overline{E^2} - \overline{E}^2 \, \rangle
= \langle \,\overline{N} ( \overline{\omega^2} - \overline{\omega}^2\,
) \rangle
+ \langle (\overline{N^2} - \overline{N}^2\, ) \overline{\omega}^2
\,\rangle
\;.
\ea
One sees that $\delta \overline{E} = 0$ for vanishing dynamical
fluctuations {\it i.e.} when $W(\zeta) = \delta (\zeta -\zeta_0)$.
Assuming again that the multiplicity distribution $P_\zeta(N)$ is
poissonian, then $\overline{N^2} - \overline{N}^2 = \overline{N}$, and
$\langle (\delta E)^2 \rangle$ reads
\be
\langle (\delta E)^2 \rangle
= \langle \,\overline{N} \: \overline{\omega^2} \rangle
= \langle \langle N \rangle \rangle \; \int d\omega \; \omega^2
P_{\rm incl}(\omega) \;,
\ee
where $P_{\rm incl}(\omega)$ is the single particle inclusive energy
distribution defined as
\be
P_{\rm incl}(\omega) \equiv {1 \over \langle \langle N \rangle \rangle} \;
\sum_N N \int d\zeta \; W(\zeta) \; P_\zeta(N) \; P_\zeta(\omega) \;.
\ee
Thus, the relative dynamical fluctuations of energy equal
\be
\bigg({\delta \overline{E} \over \langle \langle E \rangle \rangle}
\bigg)^2 =
\bigg({\Delta E \over \langle \langle E \rangle \rangle} \bigg)^2
- { \lambda \over \langle \langle N \rangle \rangle} \;,
\ee
where
\be
\lambda \equiv { \int d\omega \; \omega^2 P_{\rm incl}(\omega) \over
\Big( \int d\omega \; \omega P_{\rm incl}(\omega) \Big)^2}  \;.
\ee

\vspace{0.2cm}\noindent {\bf 21.}
In general, the statistical fluctuations are not poissonian, and
{\it a priori} their form is even not known. The dynamical fluctuations
can be then measured by means of the so--called sub--event method
\cite{Voloshin:1999yf} where one considers two different,
non-overlapping but dynamically equivalent regions of the momentum space `1'
and `2'. These can be two equal to each other non-overlapping rapidity
intervals symmetric with respect to the center--of--mass rapidity.
Let $N_{1}$ and $N_{2}$ are the numbers of hadrons (e.g. negative pions)
in these regions. There is a principal difference between the dynamical
and statistical fluctuations discussed above. The statistical event--by--event
fluctuations of $N_{1}$ and $N_{2}$ in different parts of the momentum space
are uncorrelated: $P(N_{1},N_{2}) = P_{1}(N_{1})\cdot P_{2}(N_{2})$. The
dynamical fluctuations represent, according to Eq.~(\ref{dyn}), a correlated
change of the average particle numbers $\overline{N}_{1}$ and
$\overline{N}_{2}$ with that of total entropy. Since these average values
are equal to each other, $\overline{N}_{1}=\overline{N}_{2}\equiv
\overline{N}$ (the regions `1' and `2' are dynamically equivalent),
the distributions of statistical fluctuations are also the same:
$P_{1}(N_{1})\equiv P_{\overline{N}}(N_{1})$ and $P_{2}(N_{2})\equiv
P_{\overline{N}}(N_{2})$. Therefore, the total probability for
detecting $N_1$ particles in the region `1' and $N_2$ particles in the
region `2' is
\begin{equation}\label{n1n2}
{\cal P}(N_1,N_2) = \int_{0}^{\infty}d\overline{N}
~W(\overline{N})~P_{\overline{N}}(N_1)\cdot
P_{\overline{N}}(N_2)~,
\end{equation}
and the total averaging of an observable $f(N_{1},N_{2})$ provides:
\begin{eqnarray}\label{n1n2av}
\langle\langle f(N_{1},N_{2})\rangle\rangle ~ & \equiv &~
\sum_{N_{1},N_{2}} f(N_{1},N_{2})~{\cal P}(N_{1},N_{2})~\\
& = &
~\int_{0}^{\infty}d\overline{N} ~W(\overline{N})~\sum_{N_{1},N_{2}}
f(N_{1},N_{2})~ P_{\overline{N}}(N_{1}) \cdot
P_{\overline{N}}(N_{2})~.\nonumber
\end{eqnarray}
It follows from Eq.~(\ref{n1n2av}) that
\begin{equation}\label{n1-n2}
 \frac{1}{2}~ \langle \langle (N_{1}-N_{2})^{2} \rangle \rangle
= \langle \overline{N^{2}}\rangle ~-~
\langle\overline{N}^{2}\rangle \equiv \langle (\delta N)^{2}
\rangle ~.
\end{equation}
Therefore, measuring the total fluctuations of $(N_{1}-N_{2})/2$,
one obtains the dynamically averaged statistical fluctuations in the
region `1' (equal to that in the region `2'). Subtracting $\langle(\delta
N)^{2}\rangle$ from the total fluctuations in this region,
$(\Delta N)^{2}$, one finds the dynamical part, $(\delta
\overline{N})^{2}$, of interest. Similar analysis can be performed
to get the dynamical energy fluctuations.

\vspace{0.2cm} \noindent {\bf 22.}
We have assumed that only dynamical fluctuations generated at the collision
early stage lead to the particle correlations in the final state. Of course,
it is not quite true. The effects of quantum statistics also lead to the
inter-particle correlations. However, the correlation range in the momentum
space is in this case rather small, $\Delta p \approx $ 100 MeV/$c$. The
contribution of these effects can be accounted in
$\langle (\delta N)^2 \rangle$ if the selected acceptance regions are
separated by the distance significantly larger than $\Delta p$.

\vspace{0.2cm} \noindent {\bf 23.}
There are also long range correlations which have nothing to do with
the early stage dynamical correlations and cannot be accounted in
$\langle (\delta N)^2 \rangle$ by the sub-event method described above.
In particular, there are correlations due to conservation laws. Those
can be effectively eliminated if one studies only a small part of
a whole system which is constrained by the conservation laws.

\vspace{0.2cm} \noindent {\bf 24.}
A large fraction of the final state particles comes from the decays
of various hadron resonances. The existence of resonances decaying into
at least two hadrons enlarges the final state multiplicity fluctuations.
This effect can not be eliminated by use of the sub-event method. It is
because the decay products are correlated at the scale of approximately
one rapidity unit which at the SPS energy domain is comparable to the
width of rapidity distribution. To remove bias due to resonance production
and decay, we suggest to study the fluctuations of negatively charged hadrons
as typically only one negatively charged hadron comes from a single
resonance decay.

\vspace{0.2cm} \noindent {\bf 25.}
In summary, we propose a new method to study the equation of state of
strongly interacting matter produced at the early stage of nucleus--nucleus
collisions. The method exploits the properly filtered relative fluctuations
of multiplicity and energy. Within the Statistical Model of the Early Stage
\cite{Gazdzicki:1998vd} this ratio is directly related to the fluctuations
of the early  stage entropy and energy and thus is sensitive to the EoS of
the early stage matter. We show that within the model the ratio is
a non--monotonic function of the collision energy with the maximum at the
end of the mixed phase ($\approx 60$ A$\cdot$GeV). Consequently, it can be
considered as a further signal of deconfinement phase transition.

\vspace{1cm}

We are grateful to Maciek Rybczy\' nski and Zbyszek W{\l}odarczyk for
stimulating criticism. Partial support by Bundesministerium f\"ur
Bildung und Forschung (M.~G.) and by Polish Committee of Scientific
Research under grant 2P03B04123 (M.~G. and St.~M.) is acknowledged.

\newpage

\begin{figure}
%
%
\hspace{1cm}
\epsfig{file=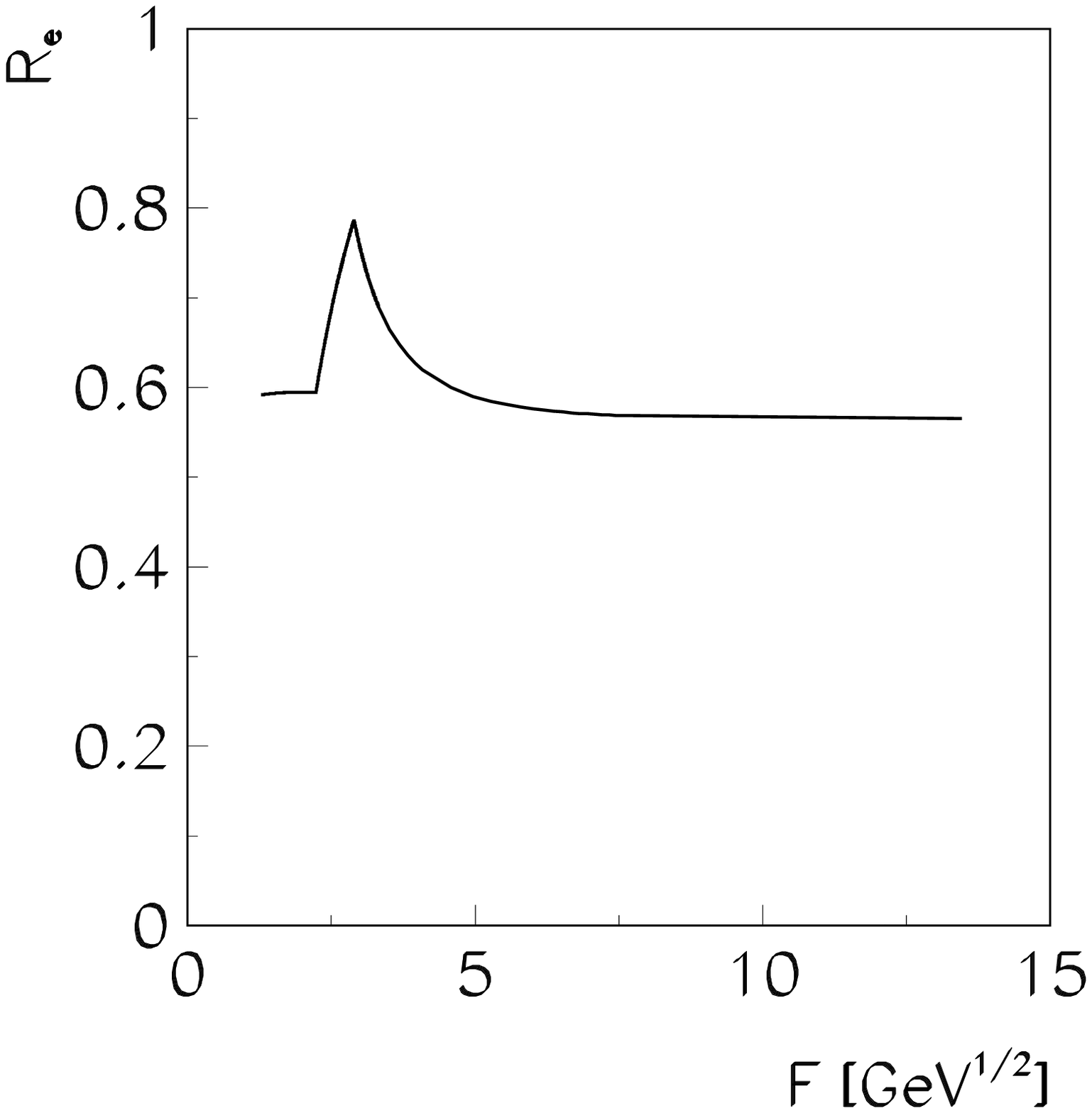,width=130mm}
\vspace{1cm}
\caption{
The dependence of $R_e$ calculated within SMES
\protect\cite{Gazdzicki:1998vd}
on the Fermi's collision energy measure
$F \equiv (\sqrt{s} - 2 m)^{3/4}/s^{1/8}$ where $\sqrt{s}$ is the c.m.s.
energy per nucleon--nucleon pair and $m$ is the nucleon mass. The `shark
fin' structure is caused by the large fluctuations in the mixed phase region.}
\label{fig1}
\end{figure}

\end{document}